\documentclass[11pt,a4paper]{article}
\usepackage{jheppub}
\usepackage{slashed}
\usepackage{graphicx}%
\usepackage{subfigure}
\usepackage{bm}
\newcommand{\beq}{\begin{equation}}
\newcommand{\eeq}{\end{equation}}

\newcommand{\be}{\begin{eqnarray}}
\newcommand{\ee}{\end{eqnarray}}
\long\def\hidestart#1\hideend{}
\title{Effect of $r$ averaging on Chiral Anomaly in Lattice QCD with Wilson
 Fermion: Finite volume and cutoff effects}

\author{Asit K. De,}

\author{A. Harindranath}

\author{and Santanu Mondal}

\affiliation{Theory Division, Saha Institute of Nuclear Physics \\
 1/AF Bidhan Nagar, Kolkata 700064, India}

\emailAdd{asitk.de@saha.ac.in}
\emailAdd{a.harindranath@saha.ac.in}
\emailAdd{santanu.mondal@saha.ac.in}
\abstract{  
We demonstrate the effectiveness of averaging over the Wilson parameter $r$
(which has been proposed earlier) in removing the cutoff effects of
naive Wilson fermions in both the anomaly term and the pseudoscalar density
 term in the flavor singlet axial Ward Takahashi identity at ${\cal O}(g^2)$
involving slowly
varying background gauge fields.  We show that it is the 
physical fermion contribution which is largely influenced  by the $r$ 
averaging. We have 
studied the possible interplay between finite size and cutoff effects
by investigating in detail naive, ${\cal O}(a)$ improved and OStm Wilson
fermion cases for a range of volumes and  lattice fermion mass ($am$). 
For naive Wilson fermions 
$r$ averaging is shown to remove the effects of the interplay.
We have shown that for the pseudoscalar density term to ${\cal O}(g^2)$
the lattice result differs from the continuum result
by exhibiting considerable $am$ dependence which appears to be a
manifestation of cutoff effects with naive Wilson fermion.
The pseudoscalar density term to ${\cal O}(g^2)$ is shown to be 
almost independent of $am$ when $r$-averaging is performed.
 }
\begin{document}
\maketitle
\flushbottom
\section{Introduction}\label{intro}
Recently we have studied \cite{anomaly1} the emergence of the  
chiral anomaly in the continuum chiral limit with 
Osterwalder-Seiler twisted mass (OStm) Wilson 
fermions \cite{osterwalder-seiler}
in lattice QCD in comparison with naive \cite{wilson1}, 
${\cal O}(a)$, and 
${\cal O}(a^2)$ improved  \cite{hamber-wu,wetzel,eguchi-kawamoto} 
Wilson fermions. Here $a$ denotes the lattice spacing.
Karsten and Smit \cite{karsten-smit}, Kerler \cite{kerler1,kerler2} and 
Seiler and Stamatescu \cite{seiler-stamatescu} have 
demonstrated some time ago 
that naive Wilson fermions on the lattice can indeed reproduce 
the chiral anomaly in the infinite volume chiral limit in 
the continuum. However
for a lattice simulation, understanding the effects of the finite volume, 
finite cutoff and nonzero fermion mass is very crucial. Possible interplay 
among  various lattice artifacts in the unimproved theory and how various 
suggested improvements help to ameliorate the situation also needs to be 
investigated. Lattice calculation of the chiral anomaly to one loop 
perturbation theory with finite lattice fermion mass ($am$) and finite 
box size ($L$) provides
an excellent laboratory to address some of these issues in a simple setting. 

Aoki \cite{aoki} proposed taking the 
average over the Wilson parameter $r= \pm 1$ for improving the scaling behavior of chiral 
condensate. David and Hamber \cite{david-hamber} also proposed similar ideas. 
For a detailed discussion of the symmetries associated with  
$r \rightarrow -r $ and their implications, specifically that the $r$ averaged
correlators are affected by only 
${\cal O}((a^2)^k), k=1,2,\ldots$, discretization 
effects,  see 
Frezzotti and Rossi \cite{frezzotti-rossi}.   
Twisted mass (tm) fermion \cite{frezzotti, sint, shindler} has an effective $r$ averaging, but at finite
lattice spacing there are isospin breaking cut off artifacts. However, in all
simulations carried out until now, ${\cal O}(a^2)$ effects have been shown to 
be under control and to disappear in the continuum limit.      
Large ${\cal O}(a^2)$ effects are only visible in the neutral pion 
mass \cite{baron}.
This  observation has been discussed in detail and explained
 theoretically in Ref.
\cite{ETMC09}.

We note in passing that since positivity property of the quark 
determinant is not  satisfied by the OStm Wilson (W) QCD action, 
numerical simulation is only possible with 
a mixed action with OStm WQCD action for the valence quarks and tm 
WQCD  action for the  sea quarks.

$r$ averaging of lattice QCD with naive Wilson fermions looks interesting.
The flavour singlet axial anomaly in this case can be studied analytically 
under certain assumptions and the effect of the $r$ averaging investigated 
quantitatively for each of the terms on the right hand side of the 
Ward Takahashi identity. In this work there is overwhelming evidence for
improvement with respect to cutoff effects, once $r$ averaging is done.
In fact the $r$ averaged result 
is much better than  ${\cal O}(a^2)$ improved Wilson fermion 
with tree level coefficients \cite{hamber-wu}. In addition there are 
interesting issues of interplay between ${\cal O}(a)$ cutoff effects and 
finite volume effects, and its removal after $r$ averaging. Contributions 
from the physical fermion and the different doubler fermions 
are also considered separately. Finally the effect of $r$ averaging on the 
pseudoscalar density (PSD) term on the right hand side of the 
Ward Takahashi identity is also studied. 
In the following paragraphs of the introduction we briefly
discuss some of these issues.

Since the emergence of anomaly is intimately tied with the removal of 
fermion doublers, the behavior of their contributions as one varies the 
parameters of the theory is of interest. Recall that as the Wilson parameter 
$r$ is set to zero, the anomaly contributions from the 16 species 
(the physical 
fermion and the 15 doublers) all cancel each other. Karsten and Smit 
showed that for non-vanishing $r$, anomaly is independent of $r$ in the limit
$a \rightarrow 0$. However, on a 
lattice with finite cutoff (practical situation), doubler masses are not 
exactly infinitely large
 and they do contribute to the anomaly depending upon the 
value of $r$. We  study the doubler contribution 
\cite{karsten-smit,hamber-wu} as a 
function of $r$ and $am$. 

We also compare  the finite volume dependence of the one loop calculation
of the anomaly with naive Wilson action and the  improved actions. The
effect of  small momentum
 behavior of the integrand as the fermion mass is lowered 
is investigated with both Periodic (P) and Anti-Periodic (AP) boundary 
conditions (BC). We find that the naive Wilson fermion has larger 
finite volume effects than the improved ones
possibly because of the interplay between the finite volume and cutoff
effects. 
Even though OStm Wilson fermion 
has a better chiral behavior as far as approach to the chiral
 limit is concerned  
than the ${\cal O}(a)$ improved Wilson fermion in the infinite volume limit
\cite{anomaly1}, 
we find that with PBC, finite volume effects are more pronounced with the 
former than with the latter. APBC improves the convergence behavior as 
$L$ becomes large for both naive and OStm Wilson fermions. The most
striking result, however, is that finite volume effects of naive Wilson 
fermions are largely removed with $r$ averaging, indicating that finite
volume effects of naive Wilson fermions observed in this calculations 
are mostly at ${\cal O}(a)$.

At the tree level, in the continuum limit, the flavour singlet
axial vector current is conserved in the chiral limit because both the 
PSD term and the contribution from the Wilson term (denoted by $\chi$
later) vanish.
To ${\cal O} (g^2)$, in the continuum limit,    
it can be  seen that the PSD term is independent of the fermion mass 
for a slowly varying background gauge field \cite{rothe} 
and produces the negative 
of the anomaly term, leading to a cancellation. 
However note that this cancellation is due to the assumption 
of slowly varying background gauge field and at this order only. 
For general gauge 
fields and to any order of $g^2$, cancellation does not occur and the flavour 
singlet axial vector Ward Takahashi identity {\em is anomalous}.   
 For ${\cal O}(a)$ improved Wilson fermions, at  ${\cal O} (g^2)$ 
 the cancellation has been demonstrated in the chiral limit for 
constant background gauge 
field by
Hamber and Wu \cite{hamber-wu}.
Our starting point is the flavour 
singlet axial vector Ward Takahashi identity on the lattice with naive and improved
Wilson fermions. 
 Our emphasis is to study the effect of
$r$ averaging on the anomaly and the PSD terms individually, 
as discussed above.
In the case of naive Wilson fermions,
 both the PSD and the anomaly terms individually 
show strong lattice mass dependence.
By using the Reisz's power counting theorem \cite{Reisz} for the PSD term to 
${\cal O}(g^2)$ for a slowly varying background gauge field we show that 
the PSD term is independent of fermion mass in the continuum limit. 
This result indicates that the observed
mass dependence with naive Wilson fermion is a pure lattice artifact which is
largely removed by $r$ averaging leading to remarkable $am$-independence.

\section{Naive, ${\cal O}(a)$ improved and OStm  Wilson Fermions}

The flavor singlet axial Ward Takahashi identity for the naive Wilson fermions on the 
Euclidean lattice reads

\begin{eqnarray}
\langle {\Delta}^{b}_{\mu} J_{5 \mu}(x) \rangle = 2m \langle 
{\overline  \psi}_x \gamma_5 \psi_x \rangle + \langle \chi_x\rangle 
\label{fsawi}
\end{eqnarray}
where $\langle {\cal O} \rangle $ denotes the functional
average of ${\cal O}$ for a background
gauge field. Explanation of other terms are as follows:
\begin{eqnarray}
{\rm The~ backward \, \, derivative,}~~~
\Delta^b_\mu f(x) &=& \frac{1}{a} \left  [ f(x) - f(x -\mu) \right ]~,
 \nonumber  \\ 
J_{5 \mu}(x) &=& \frac{1}{2} \left [ 
{\overline \psi}_x \gamma_\mu \gamma_5 U_{x,\mu} \psi_{x+\mu} 
+ {\overline \psi}_{x+\mu} \gamma_\mu \gamma_5 U^\dagger_{x \mu}
\psi_x \right ]  \nonumber \\
~~{\rm and}~~ \langle \chi_x \rangle &=& -{\rm Trace} [\gamma_5 (GW+WG)]~.
\end{eqnarray}
The Green function 
\begin{eqnarray}
G(x,y) = \langle x \mid \frac{1}{[\gamma_\mu D_\mu +W+m]} \mid y \rangle ~,
\end{eqnarray}
where
\begin{eqnarray}
[D_\mu]_{xy} &=&  \frac{1}{2a}~\left [U_{x,\mu }~\delta_{x+\mu,y}  - 
U^\dagger_{x-\mu, \mu}~\delta_{x-\mu,y} \right ]~,\nonumber \\
W_{xy} &=&  \frac{r}{2a}~\sum_\mu \left [ 2 \delta_{x,y} - 
U_{x,\mu }~\delta_{x+\mu,y}  - 
U^\dagger_{x-\mu, \mu}~\delta_{x-\mu,y} \right ]~.
\end{eqnarray}

For ease of comparisons and to facilitate discussions, here we collect the
 expressions for chiral
anomaly in the naive, ${\cal O}(a)$ improved and OStm Wilson fermions
\cite{anomaly1}.

The contribution 
to axial vector Ward Takahashi identity from the Wilson term in the {\bf unimproved} theory
\begin{eqnarray}
\langle \chi_x \rangle &=& 2 ~ g^2 ~ \epsilon_{\mu \nu \rho \lambda} ~ 
{\rm trace} ~F_{\mu \nu}(x) F_{\rho \lambda}(x)~ \frac{1}{(2 \pi)^4}\sum_p 
{\rm cos}(p_{\mu}a){\cos}(p_{\nu}a) {\cos}(p_{\rho}a)~ \nonumber \\ 
& {\hspace{.2in}}\times & W_0(p) \Big [ {\cos}(p_\lambda a) [ m + W_0(p)] 
                             - ~4~\frac{r}{a}~ {\sin}^2 (p_\lambda a) \Big ]
({\cal G}_0(p))^3 , \label{caw1} \\
&=&  \frac{g^2}{8 (\pi)^4} ~ \epsilon_{\mu \nu \rho \lambda}~ 
{\rm trace}
 ~ F_{\mu \nu}(x) F_{\rho \lambda}(x) {\tilde I}(am,r,L) \label{chiunimpw}
\end{eqnarray}
where
\begin{eqnarray}
W_{0}(p) &= &\frac{r}{a} \sum_\mu [ (1 - {\rm cos}(a p_\mu) ) ]~,
\nonumber \\
{\cal G}_0(p) &=& \left (   
\frac{1}{a^2} \sum_\mu {\sin}^2 (a p_\mu ) + (m + \frac{r}{a} \sum_\mu 
[ 1 - {\cos}(a p_\mu )])^2
\right )^{-1}
\end{eqnarray}
and trace is in color space.

Explicitly, $ \sum_p = (\frac{1}{L})^4 \sum_{n_1,n_2,n_3,n_4}$ 
where $n_1,n_2,n_3,n_4=0,1,2,3, \cdots$. In the infinite volume, chiral limit, 
${\tilde I} \rightarrow - \frac{\pi^2}{2}$ so that 
$ - \frac {2}{\pi^2} {\tilde I} = I  \rightarrow 1$. In all our plots it is 
the function $I(am,r,L)$ called anomaly integral which we have plotted.

For {\bf ${\cal O}(a)$ improved} Wilson fermion

\begin{eqnarray}
 \langle \chi_x^{I} \rangle & = &
2~g^2~ \epsilon_{\mu \nu \rho \lambda}~{\rm trace}~ 
F_{\mu \nu}(x) F_{\rho \lambda}(x) ~ \frac{1}{(2 \pi)^4}~
\sum_p \cos(ap_\mu) \cos(ap_\rho)\cos(ap_\lambda) ~
[{\cal G}^I_0(p)]^3\nonumber \\
&\hspace{.4in}&\Big [\cos(ap_\nu)[m+ W_0(p)+W^I_0(p)] 
\nonumber \\
&\hspace{.8in}& - ~4~ \frac{r}{a}~ \sin(ap_\nu)
(\sin(ap_\nu) - \frac{1}{2} \sin(2 a p_\nu))\Big ] [W_0(p)+ W^I_0(p)]
 ~.
\end{eqnarray}
Here
\begin{eqnarray}
W_{0}(p)+W^I_0(p) = \Big [   
\sum_\mu \left [\frac{r}{a} (1 - {\rm cos}(a p_\mu) )+ 
\frac{r}{4a} (-1 + {\rm cos}(2 a p_\mu) )
\right ] \Big ],~\nonumber \\
{\cal G}^I_0(p)  = \left (   
\frac{1}{a^2} \sum_\mu {\rm sin}^2 (a p_\mu ) + (m +  \sum_\mu 
[\frac{r}{a}[ 1 - {\rm cos} (ap_\mu )] +
\frac{r}{4a}[ -1 + {\rm cos}(2a p_\mu )]] )^2
\right )^{-1}. 
\end{eqnarray}

In the case of the {\bf OStm} Wilson action 
\begin{eqnarray}
\langle \chi^{os}_x \rangle &=& 2 ~ g^2 ~ \epsilon_{\mu \nu \rho \lambda} ~ 
{\rm trace}~F_{\mu \nu}(x) F_{\rho \lambda}(x)~ \frac{1}{(2 \pi)^4}~ \sum_p 
{\rm cos}(p_{\mu}a){\cos}(p_{\nu}a) {\cos}(p_{\rho}a)~ \nonumber \\ 
& \times & W_0(p) \Big [ {\cos}(p_\lambda a) W_0(p) 
                             - ~ 4~ \frac{r}{a}~ {\sin}^2 (p_\lambda a) \Big ]
({\cal G}^{os}_0(p))^3~
 \label{caosw1} 
\end{eqnarray} 

where 
\begin{eqnarray}
({\cal G}^{os})^{-1} = D^2 - m^2 - W^2 ~.
\end{eqnarray}

\section{Wilson parameter $r$ Averaging}

\begin{figure}
\begin{centering}
\includegraphics[width=4.5in]{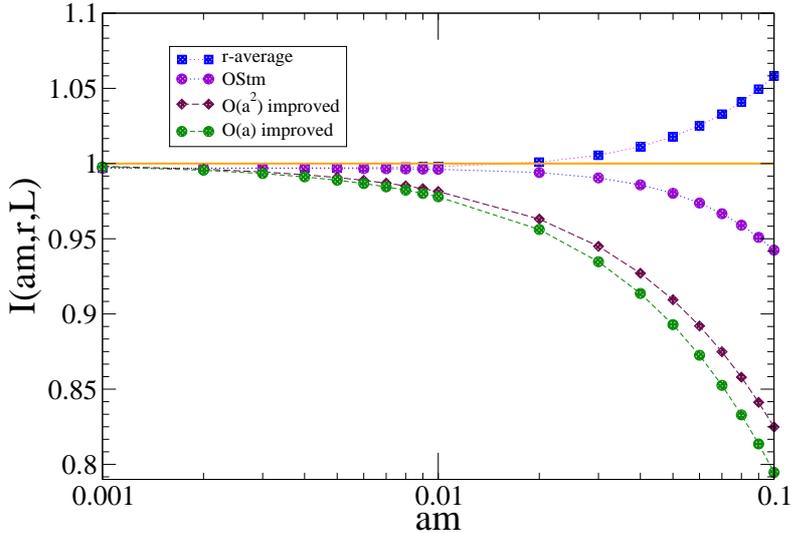}
\caption{ 
 Anomaly integral for $0.001 <am <0.1$ for $r$ averaged  
naive Wilson fermions at $L=60$ for $r=\pm 1$  
compared with 
OStm, ${\cal O}(a)$ and ${\cal O}(a^2)$ improved Wilson fermions with $r=1$.}
\label{raverage-All}
\end{centering}
\end{figure}

 In Fig. \ref{raverage-All} we compare the 
anomaly integral for $r$ averaged naive Wilson fermion
 with the OStm, ${\cal O}(a)$ and ${\cal O}(a^2)$ improved Wilson fermions.
The $r$ averaged  anomaly integral for naive Wilson fermion 
shows better chiral behaviour compared with 
 OStm Wilson fermion for $am \geq 0.01$. The cutoff effects 
are known to be proportional to even powers of $a$ for
 both $r$-averaged and OStm 
Wilson fermions but they are not the same as
 illustrated in Fig. \ref{raverage-All}. 
 Although the leading cutoff effect in $r$ averaged
naive Wilson fermion and OStm Wilson fermion is of ${\cal O}(a^2)$, 
they have much better chiral 
behaviour than the ${\cal O}(a^2)$ improved \cite{hamber-wu} Wilson fermion
which employs tree level coefficients. For further improvement 
in the latter case one needs to tune the coefficients while OStm and 
$r$-averaged Wilson fermion do not need any tuning.

In the following sections we present results of $r$ averaging for 
the anomaly integral for
different lattice volumes and $am$ and contributions to the integral from 
different 
doubler sectors.

\section{Anomaly: contribution from different regions of the Brillouin zone}
\begin{figure}
\subfigure{
\includegraphics[width=3in]{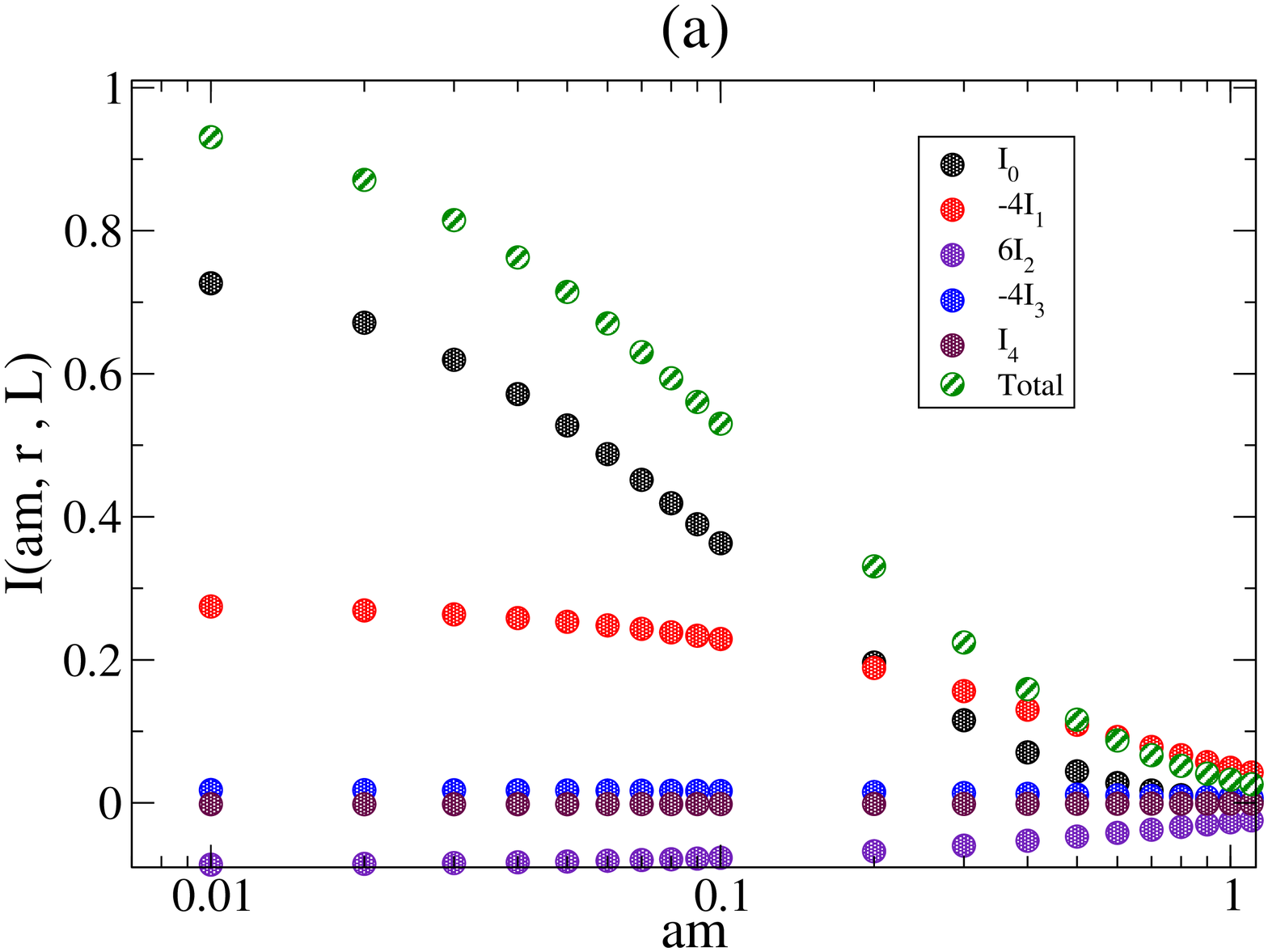}
\label{individual-ks}
}
\subfigure{
\includegraphics[width=3in]{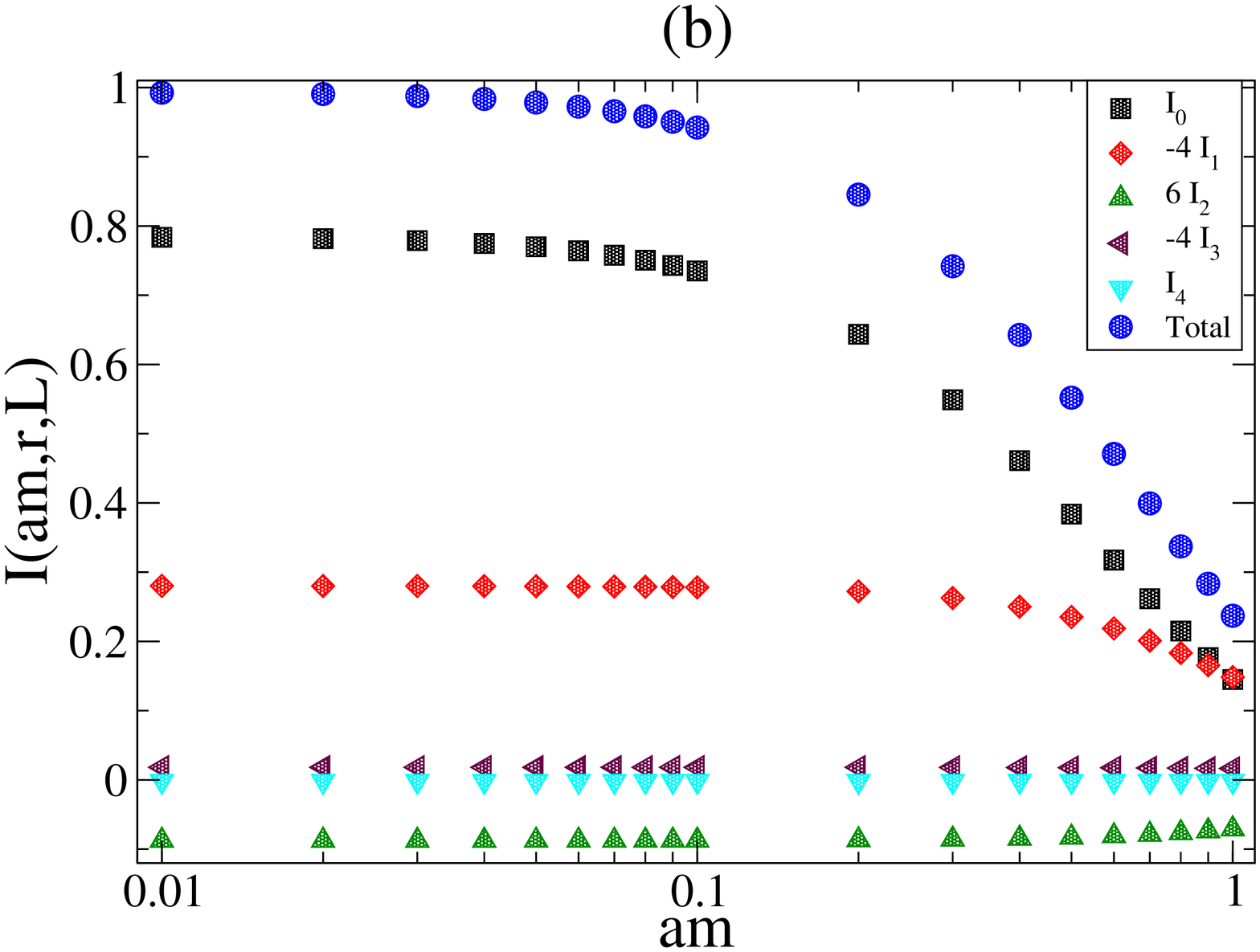}
\label{individual-ostm}
}

\subfigure{ 
\includegraphics[width=3in]{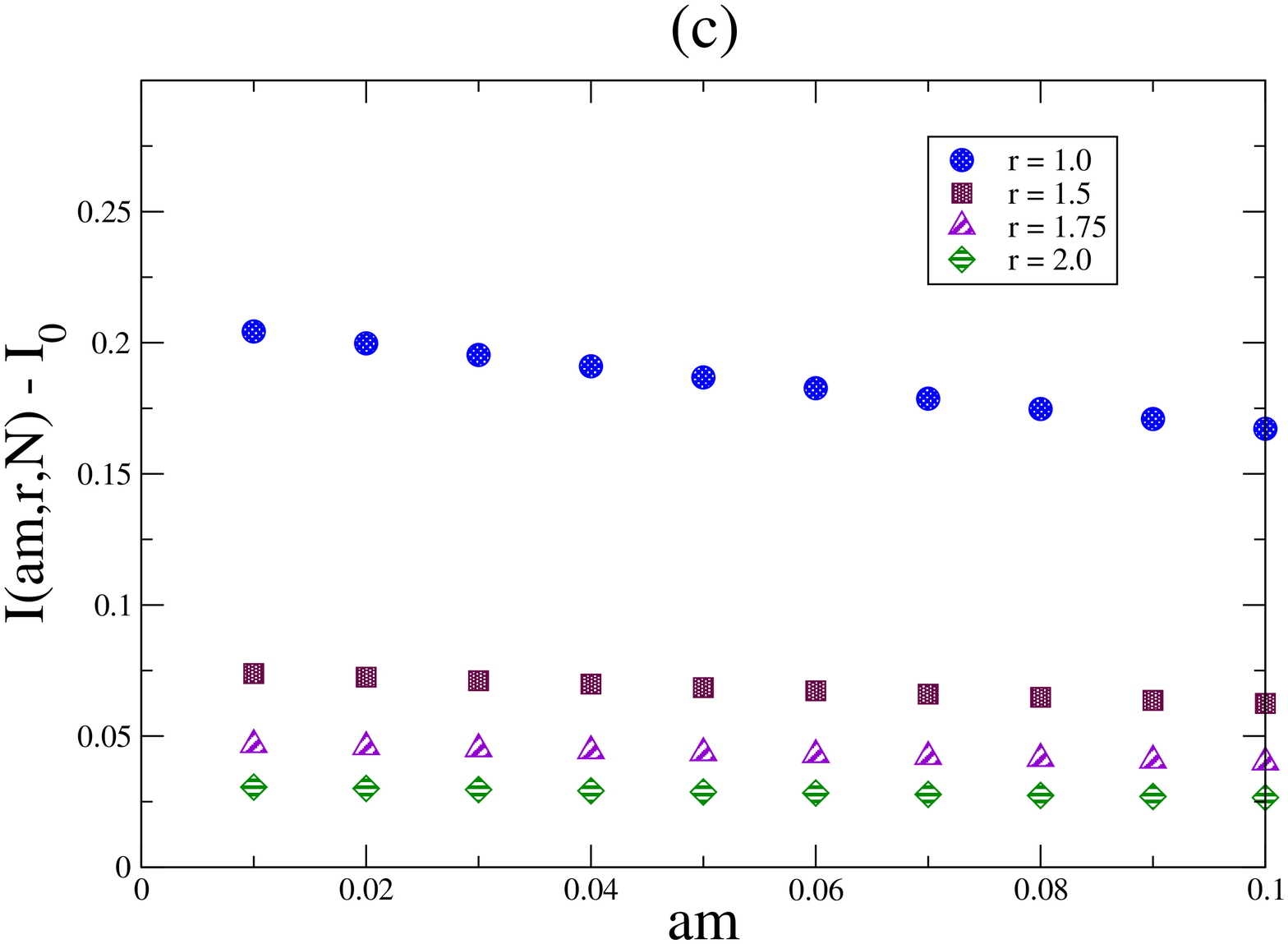}
\label{rdep-ks}
}
\subfigure{
\includegraphics[width=3in]{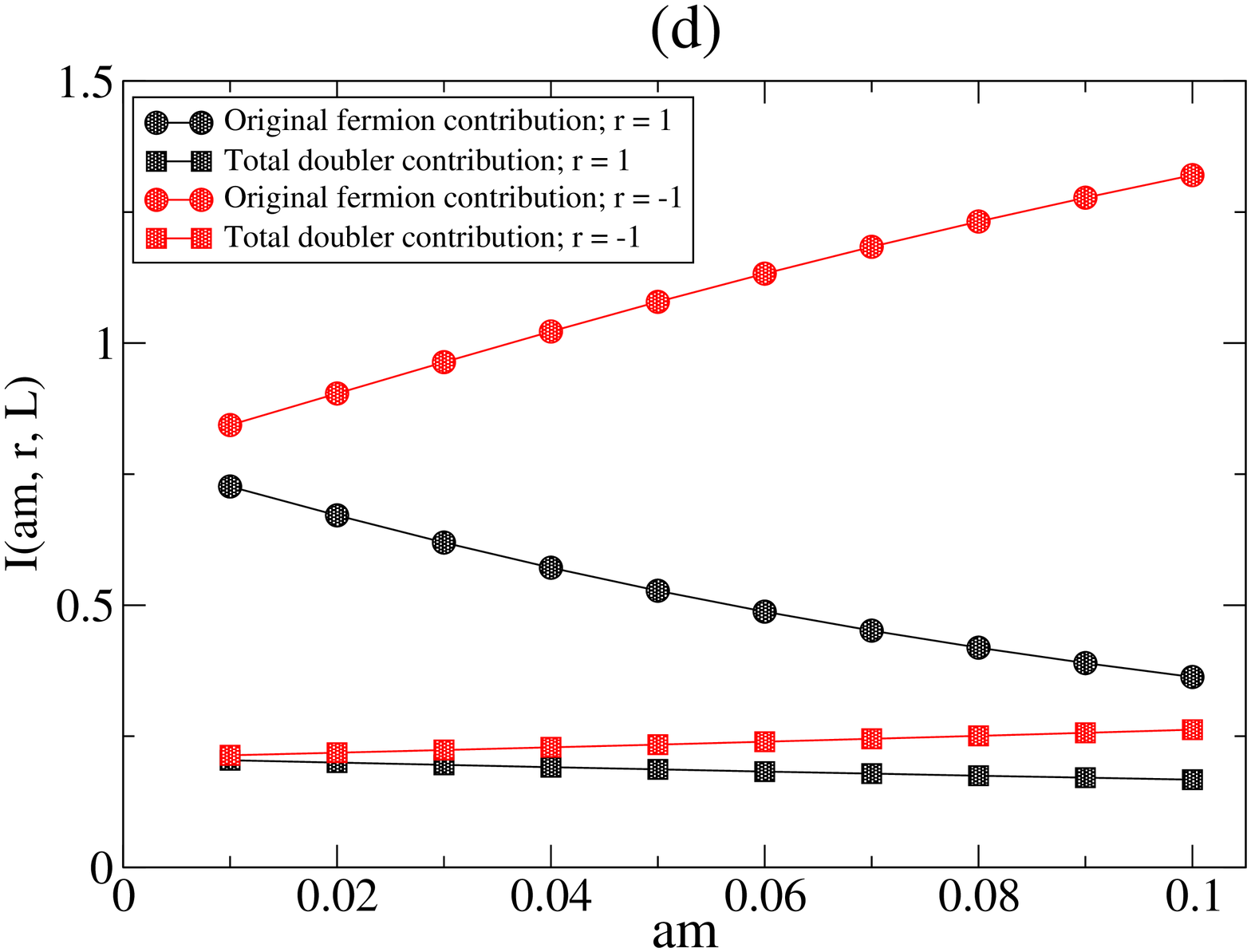}
\label{org-doubler}
}
\caption{(a) Contributions from different regions of the Brillouin zone 
to the anomaly integral for $r=1$ and $0.01 <am <1.0$ for  
naive Wilson fermions at $L=40$  compared with the 
total contribution.
(b) Contributions from different regions of the Brillouin zone 
to the anomaly integral for $r=1$ and $0.01 <am <1.0$ for  
OStm Wilson fermions at $L=40$  compared with 
total contribution.
(c) Total contribution of the doublers 
to the anomaly integral for $0.01 <am <0.1$ for 
naive Wilson fermions at $L=40$ for $1 \le  r \le 2$.
(d) Comparison of total doubler contribution and the physical fermion contribution
 for $0.01 <am <0.1$ with 
naive Wilson fermions at $L=40$ and $r=\pm 1$.
}
\end{figure}

It is instructive to separately consider the contribution to the anomaly 
integral from different regions of the Brillouin zone 
 in order to compare and 
contrast the contribution of the physical fermion and the doublers as 
a function of the different parameters. This has been studied in the 
chiral limit by Hamber and Wu \cite{hamber-wu}. However, for practical 
simulations the behavior at non-zero quark masses and nonzero
lattice spacing is more relevant and we 
study this issue in this section.   
Following Karsten and  Smit, the 
limits on the momentum sum are changed from 
$ (- \pi/a, + \pi/a) $ to 
$ (- \pi/(2a),3\pi/(2a)) $ and further  the momentum 
sum 
hypercube is divided into 16 smaller hypercubes corresponding to  
$(- \pi/(2a), +\pi/(2a)) $ and $(+ \pi/(2a), + 3 \pi/(2a))$
for each  $p_\mu, \mu=1,2,3,4 $. Thus the total anomaly contribution is 
decomposed into the contributions from the five species-types and the Anomaly 
Integral  $I=I_0-4I_1+6I_2-4I_3+I_4$.   In Figs. \ref{individual-ks} 
and \ref{individual-ostm} we compare the contributions in the region
 $0.01 <am <1.0$  
for $r= 1$ for naive Wilson fermions and OStm Wilson  fermions 
respectively.
 As expected, as the physical fermion gets lighter, its 
contribution to the anomaly is dominant compared to the doublers. Nevertheless,
doublers' contribution is not insignificant. The first doubler 
contributes the most since it has the lowest mass among the doublers. Also
note that the lowest mass doubler has more $am$ dependence in the case of
naive Wilson fermion compared to OStm fermion.  

To verify that the doublers do 
decouple when their masses become very heavy so that the sole contribution 
to the anomaly comes from the physical fermion, we study the behavior of
$I_i, i=1,2,3,4$ as the Wilson parameter $r$ is increased. We realize that 
the reflection positivity is not guaranteed when $\mid r\mid  > 1$, but 
in the 
present context we adjust r only to make the doubler masses heavy.

In Fig. \ref{rdep-ks} we present the total contribution of the doublers 
to the anomaly integral for $0.01 <am <0.1$ for 
naive Wilson fermions at $L=40$ for $1 \le  r \le 2$. As expected, the total
doubler 
contribution diminishes as $r$ increases and it is not very sensitive to $am$
in the range $0.01 \leq am \leq 0.1$. 
Fig. \ref{org-doubler} shows, for $r=\pm 1$, total doubler contribution and 
physical fermion 
contribution separately. For each $r$, total 
contributions from the 
doublers
have slight  $am$ dependence. Although for each $r$, the contribution 
from the physical fermion depends significantly on $am$, the figure clearly 
indicates that, on $r$ averaging, the $am$-dependence will cancel out.

\section{Volume Dependence}

Since numerical simulations are performed at finite volume and finite lattice 
spacing, it is of interest to study the finite volume dependence of the 
anomaly integral as a function of the quark mass. Since finite size artifacts 
are different for the three fermion actions considered in this work, and 
finite volume  and finite cutoff (1/$a$) effects may interfere with each other,
the anomaly integral provides us an opportunity to explore this issue in this 
simple setting.     
\begin{figure}
\subfigure{
\includegraphics[width=3in]{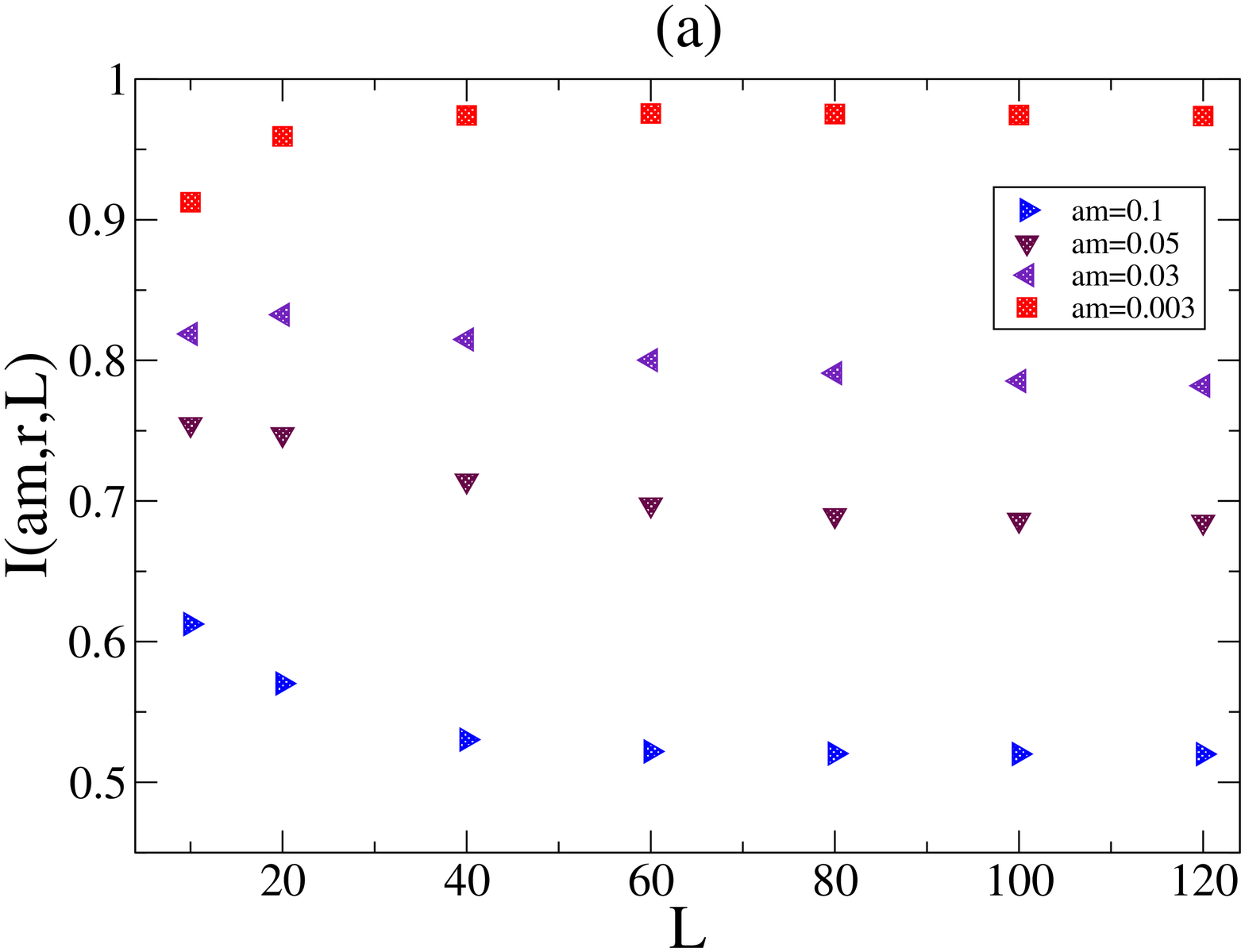}
\label{ks-ldep}
}
\subfigure{
\includegraphics[width=3in]{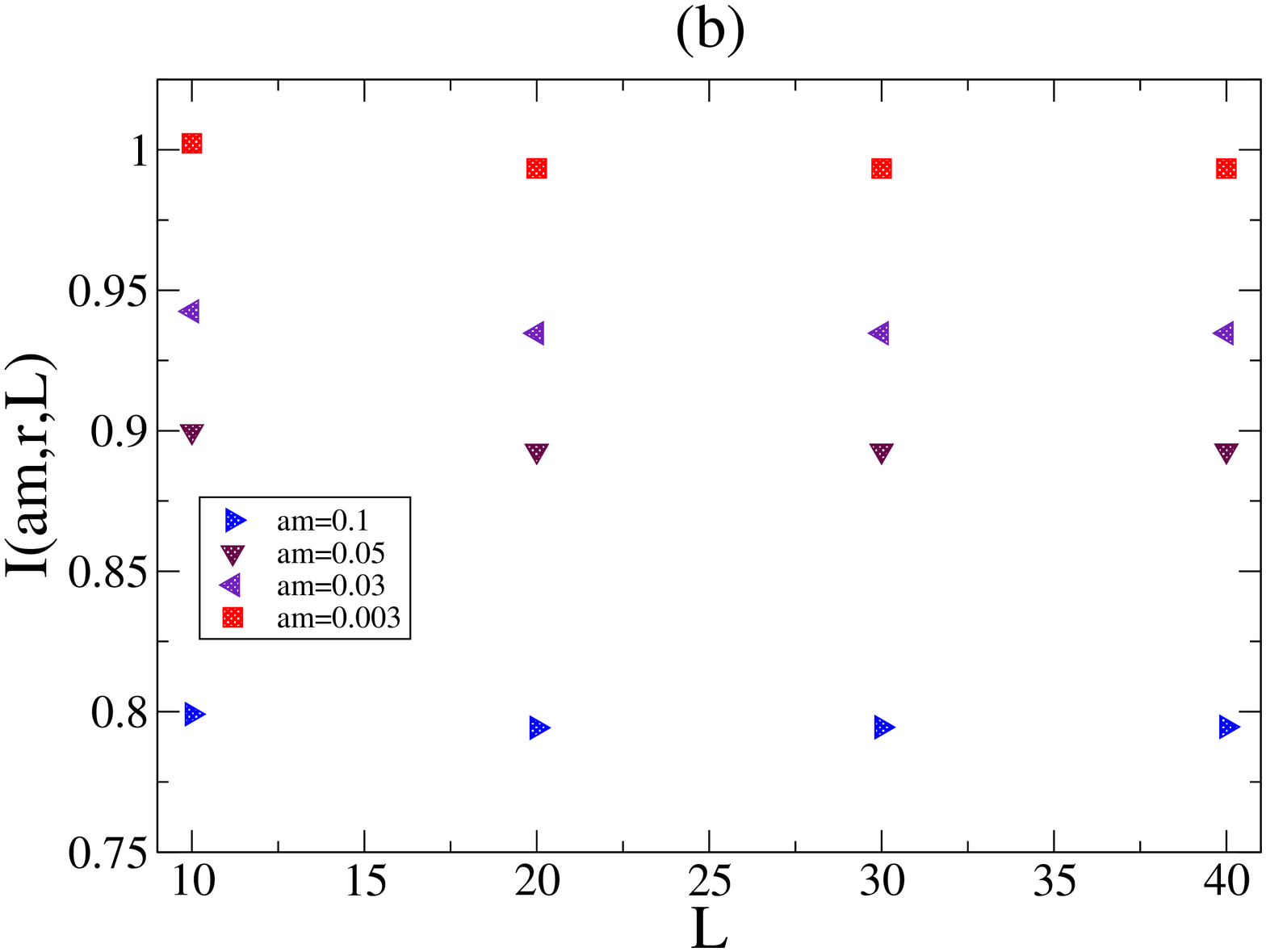}
\label{hw-ldep}
}
\subfigure{
\includegraphics[width=3in]{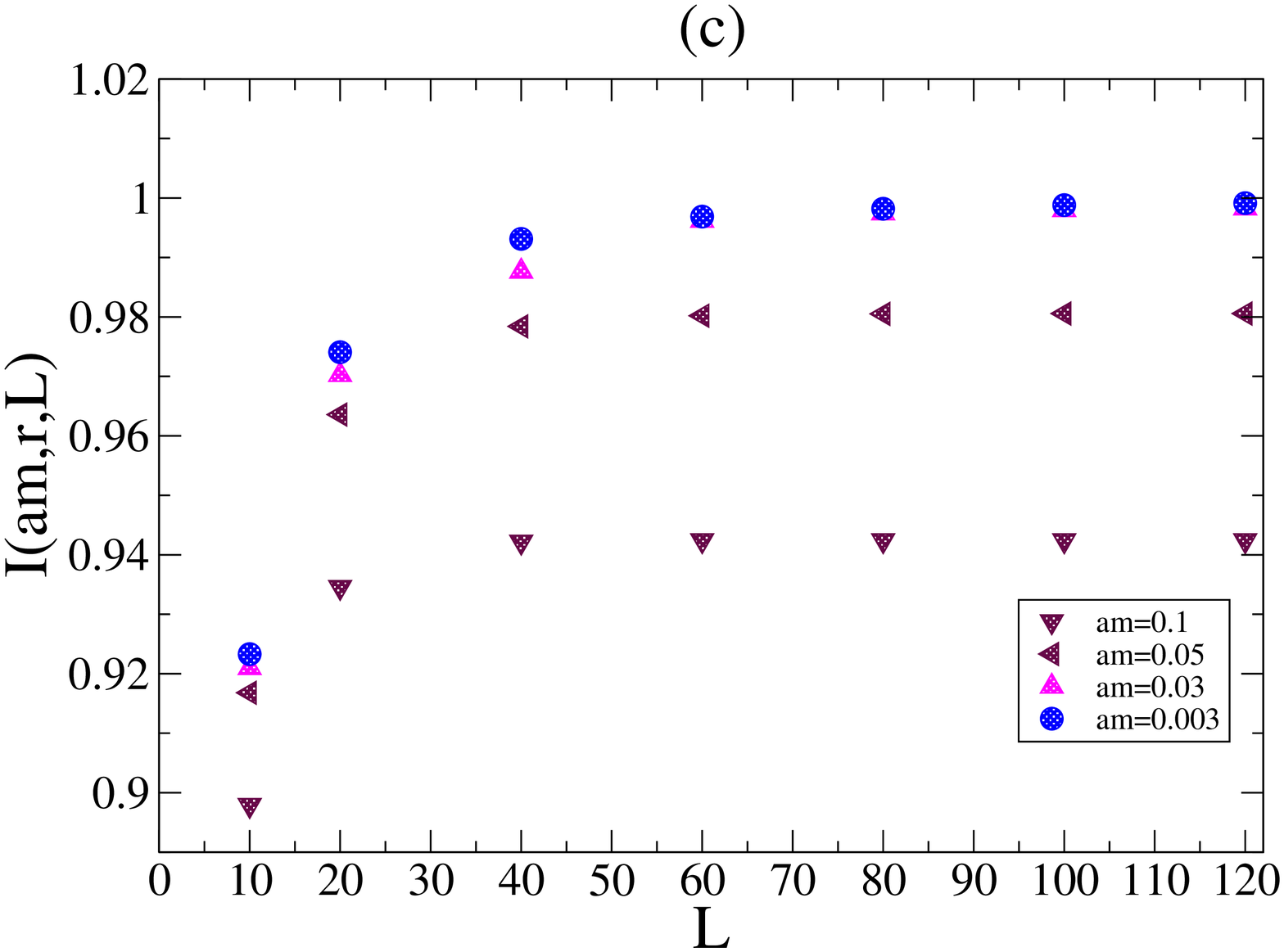}
\label{ostm-ldep}
}
\subfigure{
\includegraphics[width=3in]{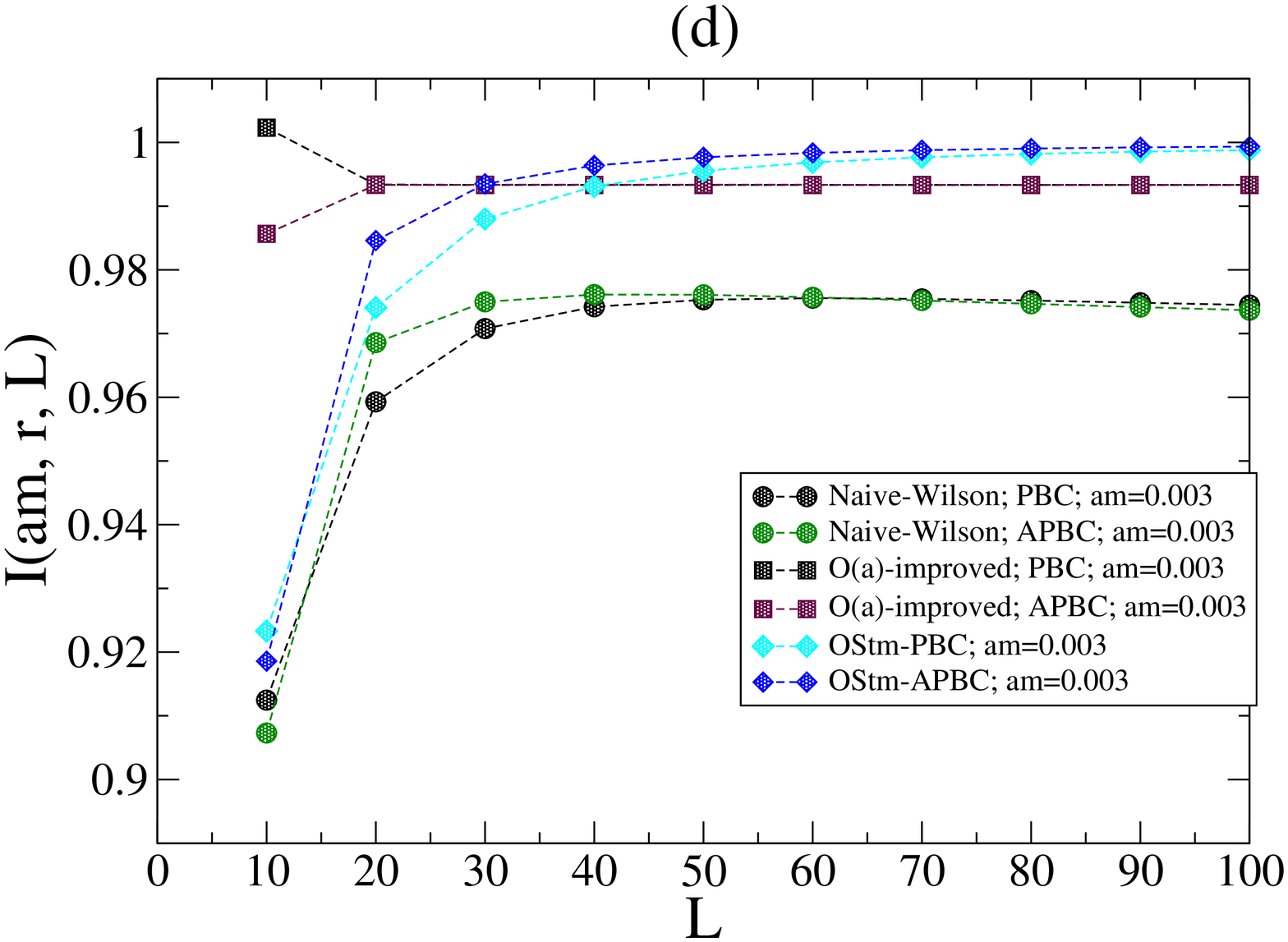}
\label{ks-hw-ostm-l-m-dep}
}
\caption
{
(a) Volume dependence of  
the anomaly integral for $r=1$ and $0.01 \le am \le 0.1$ for 
naive Wilson fermions with PBC.
(b) Volume dependence of  
the anomaly integral for $r=1$ and $0.003 \le am  \le 0.1$ for 
${\cal O}(a)$ improved Wilson fermions with PBC.
(c) Volume dependence of  
the anomaly integral for $r=1$ and $0.003 \le am \le 0.1$ for 
OStm Wilson fermions with PBC.
(d) Volume dependence of  
the anomaly integral for $am = 0.003$ and
$ 10 \le L \le 120$ 
for naive Wilson, ${\cal O}(a)$ improved Wilson and 
OStm Wilson fermions with PBC and APBC.
}
\end{figure}

\begin{figure}
\begin{centering}
\includegraphics[width=4.5in]{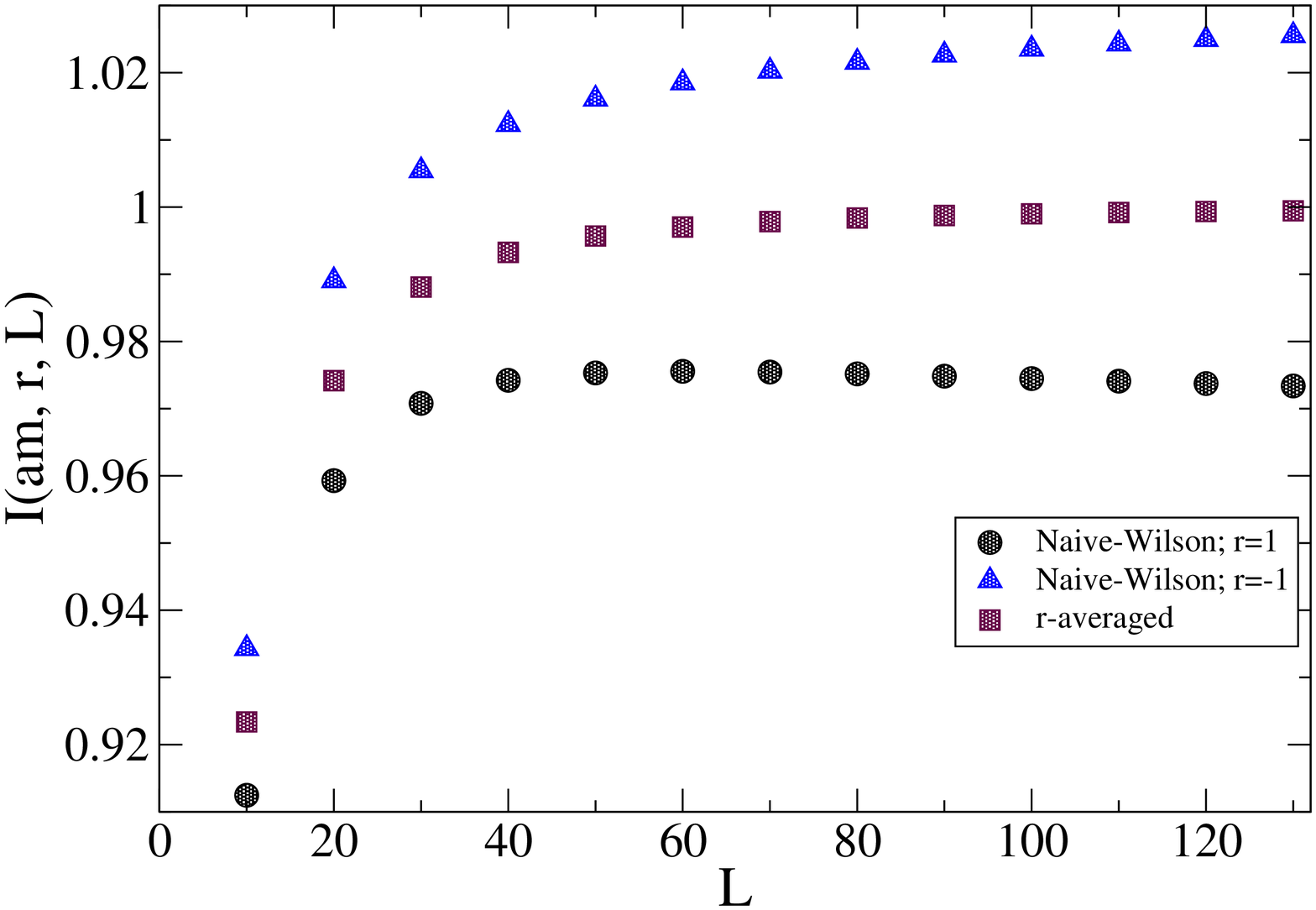}
\caption{Volume dependence of  
the anomaly integral for $am = 0.003$ and
$ 10 \le L \le 130$ 
 for $r=\pm 1$ 
and $r$ averaged for naive Wilson with PBC.}
\label{ks-r-avg-l-m-dep}
\end{centering}
\end{figure}

First we discuss the case of PBC in all four 
directions. 
In Fig. \ref{ks-ldep} volume dependence of  
the anomaly integral for $0.01 \le am \le 0.1$ for 
naive Wilson fermion for $r=1$ and PBC
 is presented.  We note that the behavior 
qualitatively changes as $am$ changes from large  to small. Moreover, 
irrespective of the quark mass, approximate $L$ independence is achieved 
only for 
$L\gtrsim 40$. In Fig. \ref{hw-ldep} the corresponding behavior for $O(a)$ 
improved Wilson fermion is shown. Here the $L$ dependence is qualitatively 
the same  across the range of $am$, and furthermore,  $L$ independence 
is achieved 
for $L>20$.  Volume dependence of  
the anomaly integral for OStm Wilson fermion is presented in 
Fig. \ref{ostm-ldep}. Here again the $L$ dependence is qualitatively 
the same  across the mass range but $L$ independence is achieved 
only for 
$L > 50$. Nevertheless the infinte volume result for OStm fermion is closer 
to the continuum  compared to ${\cal O}(a)$ improved result.  
The small $L$ behavior is different for the three types of fermions
considered here.

We have checked the sensitivity with respect to boundary condition of the 
volume dependence of the anomaly integral for the three actions in 
Fig. \ref{ks-hw-ostm-l-m-dep}.  For the $O(a)$ improved Wilson fermion, there 
is sensitivity to boundary condition only for very small $L$. On the other hand
both for the naive and OStm Wilson fermion, the APBC improves the 
convergence of the anomaly integral with respect to $L$.  

In Figs. \ref{ks-ldep}, \ref{hw-ldep}, \ref{ostm-ldep} and  
\ref{ks-hw-ostm-l-m-dep}
we have seen some interplay between the finite volume and the cutoff
effects for naive Wilson fermion. The fact that this interplay is absent 
for ${\cal O}(a)$ improved Wilson fermion indicates that this is largely an 
${\cal O}(a)$
artifact.
Finally  Fig. \ref{ks-r-avg-l-m-dep} shows that the convergence of the
anomaly integral with respect to $L$ is vastly improved by $r$ averaging, 
since it removes the cutoff effects to a great extent.  
\section{Pseudoscalar density term versus the anomaly term}


After the detailed analysis of the anomaly integral with respect to the $am$, 
$r$ and $L$ dependence, we revisit the flavor singlet axial Ward Takahashi 
identity, Eq. (\ref{fsawi}). 
To ${\cal O}(g^2)$, for slowly varying background gauge fields
 the contribution from the PSD term is 

\begin{eqnarray}
2 m ~\langle {\bar \psi}_x \gamma_5 \psi_x \rangle &=& 2 ~ g^2 
~ \epsilon_{\mu \nu \rho \lambda} ~ {\rm trace} ~
F_{\mu \nu}(x) F_{\rho \lambda}(x)~ \frac{1}{(2 \pi)^4}\sum_p 
{\rm cos}(p_{\mu}a){\cos}(p_{\nu}a) {\cos}(p_{\rho}a)~ \nonumber \\ 
& {\hspace{.2in}}\times & m \Big [ {\cos}(p_\lambda a) [ m + W_0(p)] 
                             - ~ 4~\frac{r}{a}~ {\sin}^2 (p_\lambda a) \Big ]
({\cal G}_0(p))^3 , \\
&=&  \frac{g^2}{8 (\pi)^4} ~ \epsilon_{\mu \nu \rho \lambda}~ 
{\rm trace} ~ F_{\mu \nu}(x) F_{\rho \lambda}(x) 
{\tilde I}_m(am,r,L) \label{munimpw}
\end{eqnarray}
where 
\begin{eqnarray}
{\tilde I}_m(am,r,L) &=&
\sum_p 
{\rm cos}(p_{\mu}a){\cos}(p_{\nu}a) {\cos}(p_{\rho}a)~ \nonumber \\ 
& {\hspace{.2in}}\times & m \Big [ {\cos}(p_\lambda a) [ m + W_0(p)] 
                             - ~ 4~\frac{r}{a}~ {\sin}^2 (p_\lambda a) \Big ]
({\cal G}_0(p))^3~. \label{mterm}
\end{eqnarray}  

First by using Reisz power counting theorem \cite{Reisz} we show that
 the PSD term to ${\cal O}(g^2)$ is independent of $m$ in the infinite volume 
continuum limit.
In the limit $L \rightarrow \infty $ 
\begin{eqnarray}
{\tilde I}_m(am,r) 
=\int^{\frac{\pi}{a}}_{-\frac{\pi}{a}}~d^4p~
\frac{N(m,p,a)}{D(m,p,a)}\label{Reisz1}
\end{eqnarray}
where
\begin{eqnarray}
N(m,p,a)&=&{\rm cos}(p_{\mu}a){\cos}(p_{\nu}a) {\cos}(p_{\rho}a)
\times  m \Big [ {\cos}(p_\lambda a) [ m + W_0(p)]
 - ~ 4~\frac{r}{a}~ {\sin}^2 (p_\lambda a) \Big ]\nonumber\\
&=&{\rm cos}(p_{\mu}a){\cos}(p_{\nu}a) {\cos}(p_{\rho}a)           
\times  m \Big [ {\cos}(p_\lambda a) [ m + \frac{r}{a} 
\sum_\mu[ 1 -{\cos}(a p_\mu )]]
 - ~ 4~\frac{r}{a}~ {\sin}^2 (p_\lambda a) \Big ]\nonumber\\
D(m,p,a)&=&{\cal G}_0(p)^{-3}=\left (   
\frac{1}{a^2} \sum_\mu {\sin}^2 (a p_\mu ) + (m + \frac{r}{a} \sum_\mu
[ 1 - {\cos}(a p_\mu )])^2
\right )^3\nonumber
\end{eqnarray} 
To find lattice degree of divergences (LDD) 
 of $N(m,p,a)$ and $D(m,p,a)$ we scale
$p$ and $a$ in the following way such that $\hat{p}=pa$ remains fixed,
\begin{eqnarray}
p&\rightarrow &\lambda p\nonumber\\
a&\rightarrow& a/\lambda \nonumber.
\end{eqnarray}
Then,
\begin{eqnarray}
N(m,p\lambda,a/\lambda)&=& \frac{r}{a}\Big[m\sum_\mu[ 1 -{\cos}(a p_\mu
)]-~4~{\sin}^2 (p_\lambda a)\Big]~\lambda + {\cal O}(\lambda^0)\nonumber\\
D(m,p\lambda,a/\lambda)&=& \Big [ \frac{1}{a^6}[\sum_\mu {\sin}^2 (a p_\mu
)]^3+\frac{r^6}{a^6}[\sum_\mu[ 1 -{\cos}(a p_\mu )]]^6
+\frac{3r^2}{a^6}[\sum_\mu
{\sin}^2 (a p_\mu
)]^2 [\sum_\mu[ 1 -{\cos}(a p_\mu)]]^2\nonumber\\
&+&\frac{3r^4}{a^6}[\sum_\mu{\sin}^2 (a
p_\mu)][\sum_\mu[ 1 -{\cos}(a p_\mu)]]^4\Big ]~\lambda ^6
+{\cal O}(\lambda ^5) \nonumber
\end{eqnarray}

Then according to Reisz, LDD of $N$ ($degr_{\hat{p}}N$) and LDD of 
$D$ ($degr_{\hat{p}}D$) are given by,
\begin{eqnarray}
degr_{\hat{p}}N=1\nonumber\\
degr_{\hat{p}}D=6\nonumber
 \end{eqnarray}
So, LDD of the integral in Eq. (\ref{Reisz1}) is 
$$4+1-6=-1<0$$
Then according to the Reisz power counting theorem, continuum limit of 
${\tilde I}_m(am,r)$ is
\begin{eqnarray}
 &\int ^{\infty}_{-\infty}&~d^4p~ 
~\lim_{a\rightarrow 0}\Big[ {\rm cos}(p_{\mu}a){\cos}(p_{\nu}a) {\cos}(p_{\rho}a)
~\times ~ m \Big [ {\cos}(p_\lambda a) [ m + W_0(p)]
                             - ~ 4~\frac{r}{a}~ {\sin}^2 (p_\lambda a) \Big ]
({\cal G}_0(p))^3\Big]~\nonumber\\
&=& \int ^{\infty}_{-\infty}~d^4p~\frac{m^2}{(p^2+m^2)^3}
=\frac{\pi^2}{2}.\nonumber 
\end{eqnarray}
\begin{eqnarray}
{\tilde I}_m(am,r,L) = - \frac{2}{\pi^2}{I}_m(am,r,L)~.
 \end{eqnarray} 
$$\lim_{a\rightarrow 0,~ L\rightarrow \infty}{I}_m(am,r,L)=-1$$ and hence
$\lim_{a\rightarrow 0,~ L \rightarrow \infty}{I}_m(am,r,L)=-1$ is
independent of $m$.

\begin{figure}
\subfigure{
\includegraphics[width=3in]{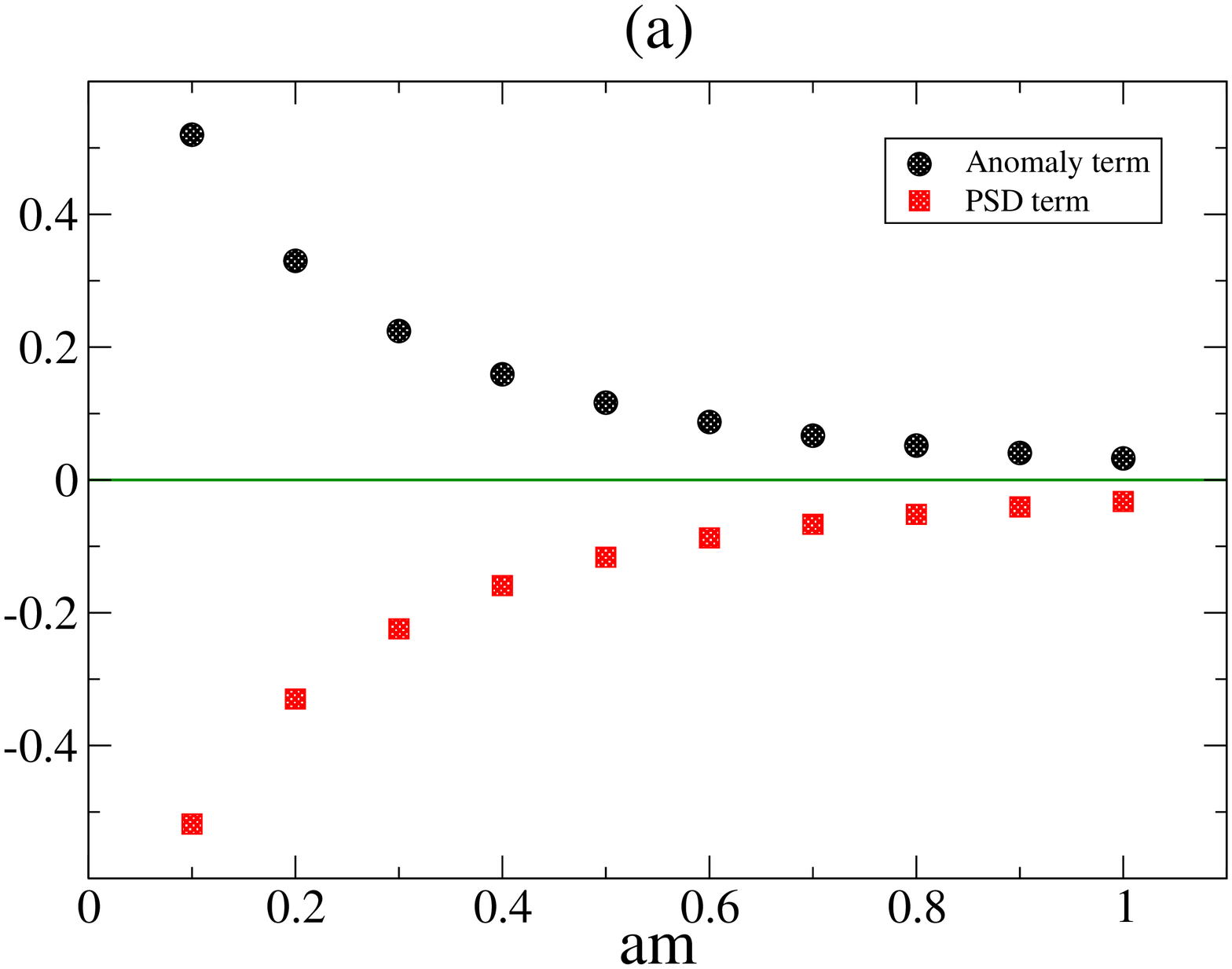}
\label{mtamdep}
}
\subfigure{
\includegraphics[width=3in]{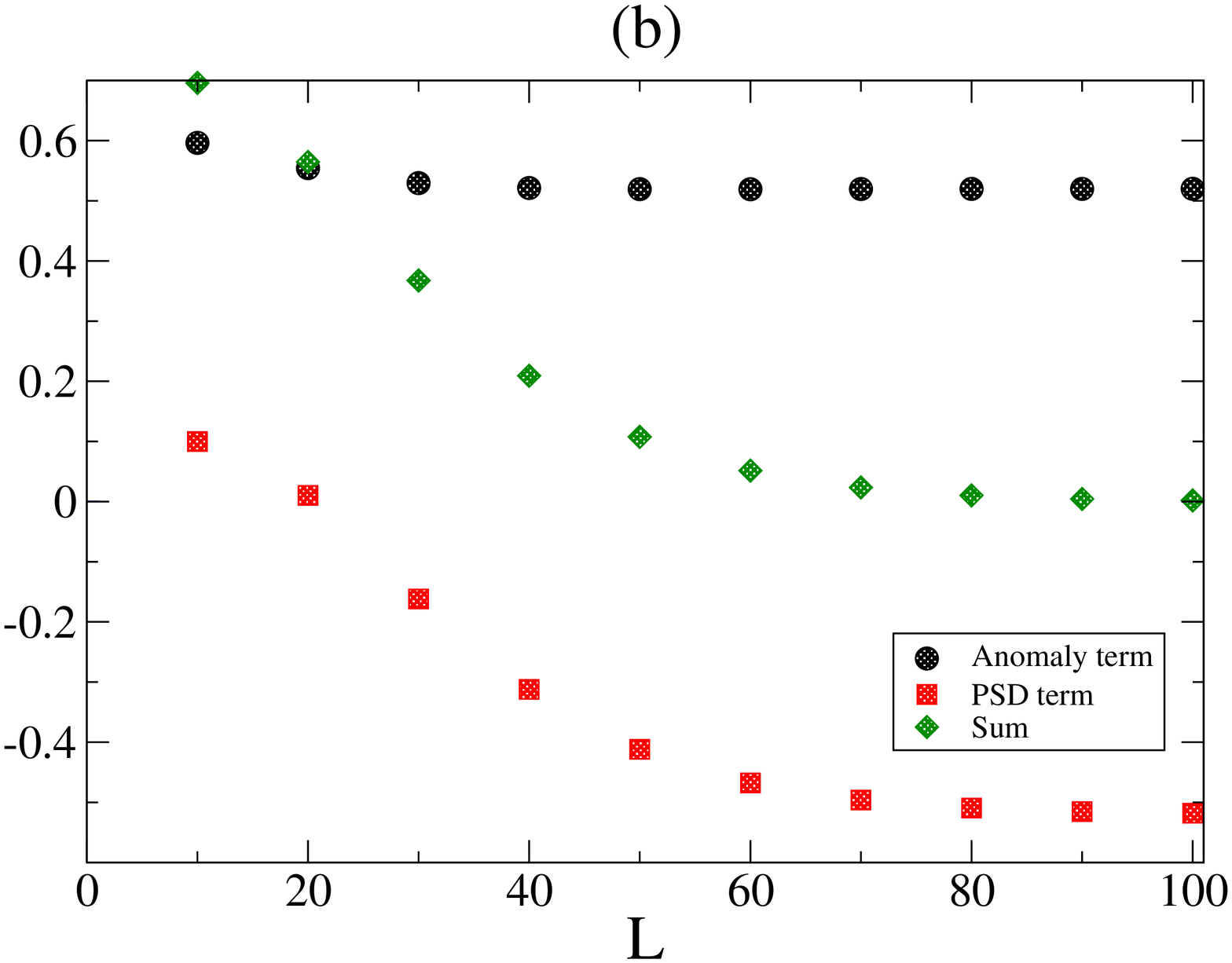}
\label{mvaldep}
}
\begin{centering}
\subfigure{
\includegraphics[width=4.5in]{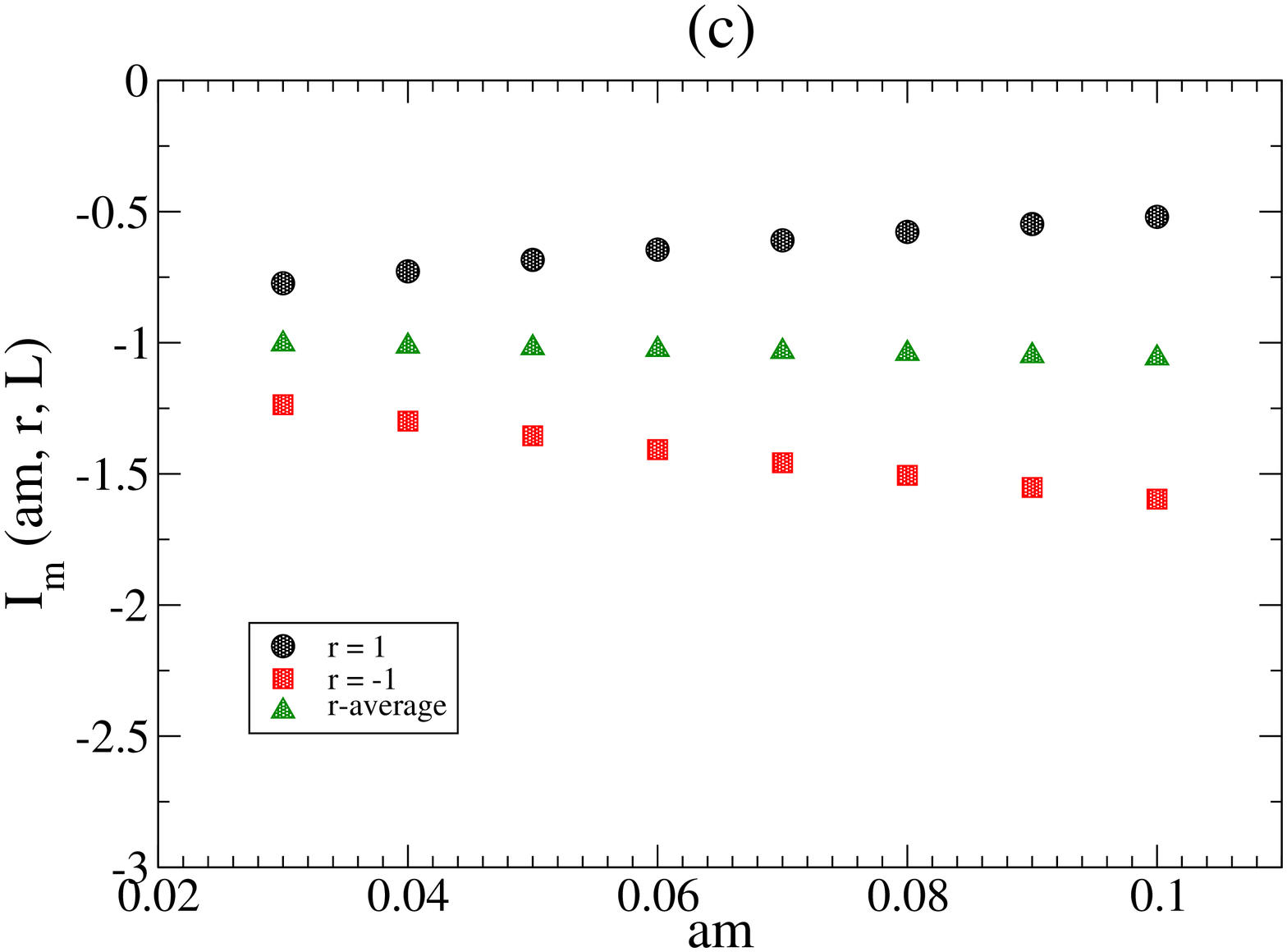}
\label{mravg}
}
\end{centering}
\caption
{
(a) The $am$ dependence of the anomaly term and the  PSD term 
for $L=100$ at ${\cal O}(g^2)$.
(b) 
The $L$ dependence of the anomaly term, the PSD term and the sum 
for $am=0.1$ ${\cal O}(g^2)$.
(c)
The PSD term for $r = \pm 1$ and the average, for $0.03\leq am \leq 0.2$ 
and $L=300$ ${\cal O}(g^2)$.
}
\end{figure}

For the PSD term 
 when $am$ 
decreases one needs to go to larger value of $L$ 
compared to the anomaly term to achieve the infinite volume
limit. It is important to note that as long as $am $ is nonzero, there is 
no infrared singularity. 
   This is borne out by numerical calculations as shown in Fig. \ref{mvaldep}. 
Furthermore we see that the lattice result differs from the continuum 
result by exhibiting considerable $am$ dependence 
(see Fig. \ref{mtamdep}) which 
appears to
be a manifestation of cutoff effects with naive Wilson fermion.
Nevertheless we see that the two terms, namely, the
 anomaly term and the PSD term to ${\cal O}(g^2)$ cancel each 
other independent of the
value of $am$ for large enough volumes.       
To check whether $r$ averaging can remove the cutoff effects we study the 
$am$ dependence of the $r$ averaged PSD term.
Fig. \ref{mravg} shows that the PSD term is almost independent of $am$ 
(for $am<=.1$)
as in the continuum when $r$-averaging is performed.

\section{Conclusions}
In this work we have addressed four issues related to the one loop lattice 
calculation of chiral anomaly in the case of Wilson-like fermions.

Some general remarks are in order. The weak coupling perturbative analysis 
may not be sensitive to some lattice artifacts that may be present in 
numerical simulations. On the other hand, some of the cutoff effects found in 
the perturbative analysis  especially for the naive Wilson fermions 
could  be removed by smoothing of gauge fields and smearing of hadronic
operators in
numerical simulations. 
  
We have used one loop lattice calculation of chiral anomaly to study 
quantitatively the effects of $r$ averaging which has been proposed before 
to achieve better scaling behavior of observables with Wilson fermions. 
We find that $r$ averaging of naive Wilson fermion has a slightly better 
approach to continuum chiral limit compared to OStm Wilson fermions.
We show that in this case $r$
averaged result is much better than 
${\cal O}(a)$ and even than ${\cal O}(a^2)$ improved Wilson fermion
with tree level coefficients. 
It is the 
physical fermion contribution which is largely influenced  by the $r$ 
averaging as seen in Fig. \ref{org-doubler}.

We have studied the doubler contributions as a function of $r$ and lattice
fermion mass $am$. We have verified that the doubler contribution decreases as 
$r$ increases and it is not very sensitive to the lattice fermion mass.

Next, we have 
studied the possible interplay between finite size and cutoff effects
by investigating in detail naive, ${\cal O}(a)$ improved and OStm Wilson
fermion cases for a range of volumes and $am$. We find that the major 
contribution to the interplay between finite volume and cutoff effects is 
at the ${\cal O}(a)$ level for naive Wilson fermion.

Lastly we have studied the relative roles played by the PSD term and the 
anomaly term in the flavor singlet axial Ward Takahashi identity 
for naive Wilson fermion on the lattice at
one loop level for slowly varying background gauge fields.
First by using Reisz power counting theorem we have shown that the PSD term
to ${\cal O}(g^2)$ is independent of $m$ and cancels the anomaly 
term in infinite volume 
continuum limit.
The lattice result for PSD term differs from the continuum 
result by exhibiting considerable $am$ dependence for $r= +1$ or $r=-1$
  (see Figs. 
\ref{mtamdep} and \ref{mravg})
 which appears to
be a manifestation of cutoff effects with naive Wilson fermion.
Nevertheless we find that the two terms, namely, the
 anomaly term and the PSD term to ${\cal O}(g^2)$ 
cancel each other independent of the
value of $am$ for $r= +1$ or $r=-1$ for large enough volumes.       
To verify whether $r$ averaging can remove the cutoff effects we investigate
 the 
$am$ dependence of the $r$ averaged PSD term.
Fig. \ref{mravg} shows that the PSD term is almost independent of $am$
 when $r$-averaging is performed.
 
In the light of the results 
obtained above, it is tempting to try out $r$-averaging in lattice
QCD simulations with naive Wilson fermions. However, problems related with
unphysical UV fluctuations may potentially        
make the simulation difficult. Smearing of gauge links in the molecular 
dynamics trajectories has been used successfully to damp these unphysical   
fluctuations \cite{capitani, duerr}. 



\end{document}